\documentclass[pra,twocolumn,superscriptaddress,showpacs,preprintnumbers,amsmath,amssymb]{revtex4}
\usepackage{}

\usepackage{color}
\usepackage[colorlinks=true]{hyperref}
\hypersetup{colorlinks=true,linkcolor=blue,citecolor=blue,urlcolor=blue}
\usepackage{bm}
\usepackage{bbm}
\usepackage{amssymb}
\usepackage{amsfonts}
\usepackage{epsfig,graphicx}
\usepackage{amstext}
\usepackage{amsmath}
\usepackage{graphicx}
\usepackage{times}
\usepackage{txfonts}
\usepackage{dcolumn}
\usepackage{dsfont}

\newcommand{\iden}{\mathds{1}}
\newcommand{\tr}{\mathrm{tr}}

\newcommand{\etal}{\textit{et al. }}

\begin{document}

\title{Sequential sharing of two-qudit entanglement based on the entropic uncertainty relation}

\author{Ming-Liang Hu}
\email{mingliang0301@163.com}
\address{School of Science, Xi'an University of Posts and Telecommunications, Xi'an 710121, China}

\author{Heng Fan}
\email{hfan@iphy.ac.cn}
\affiliation{Institute of Physics, Chinese Academy of Sciences, Beijing 100190, China}
\affiliation{CAS Center for Excellence in Topological Quantum Computation, University of Chinese Academy of Sciences, Beijing 100190, China}
\affiliation{Beijing Academy of Quantum Information Sciences, Beijing 100193, China}

\begin{abstract}
Entanglement and uncertainty relation are two focuses of quantum theory.
We relate entanglement sharing to the entropic uncertainty relation in a
$(d\times d)$-dimensional system via weak measurements with different
pointers. We consider both the scenarios of one-sided sequential
measurements in which the entangled pair is distributed to multiple
Alices and one Bob and two-sided sequential measurements in which the
entangled pair is distributed to multiple Alices and Bobs. It is found
that the maximum number of observers sharing the entanglement strongly
depends on the measurement scenarios, the pointer states of the apparatus,
and the local dimension $d$ of each subsystem, while the required minimum
measurement precision to achieve entanglement sharing decreases to its
asymptotic value with the increase of $d$. The maximum number of observers
remain unaltered even when the state is not maximally entangled but has
strong-enough entanglement.
\end{abstract}


\maketitle

\section{Introduction} \label{sec:1}
Quantum entanglement is a central topic of quantum theory, and the
distant parties sharing an entangled quantum state can establish
strong correlations that have no classical analog \cite{QE}. Aside
from its fundamental significance, it is also an essential physical
resource for various quantum information processing tasks, such as
quantum teleportation \cite{QT}, dense coding \cite{DC}, remote state
preparation \cite{RSP}, and quantum key distribution \cite{QKD}. It
is also deeply connected to other characteristics of quantum states,
including Bell nonlocality \cite{Bell1,Bell2}, Einstein-Podolsky-Rosen
(EPR) steering \cite{steer1,steer2}, quantum discord \cite{QD}, and
quantum coherence \cite{Plenio,Hu}.

Since quantum correlations (nonlocality, entanglement, EPR steering, etc.)
are vital physical resources, starting from the perspective of
quantum information processing, it would be desirable to demonstrate
these characteristics between separated observers who perform local
measurements on a shared quantum state. The usual projective (strong)
measurement induces collapse of the initial state into one of the
eigenbases of the measured observable, hence in this case the quantum
correlations can be shared by not more than two observers. Silva \etal
considered instead a different scenario in which  one Alice and
multiple Bobs (say, Bob$_1$, Bob$_2$, etc.) share an entangled pair,
with the middle Bobs (Alice and the last Bob) performing weak
(projective) measurements \cite{shareBT1}. In this context, they
showed that Bell nonlocality could be sequentially shared in the
sense that a simultaneous violation of Clauser-Horne-Shimony-Holt
(CHSH) inequalities between Alice-Bob$_1$ and Alice-Bob$_2$ was
observed. Subsequently, a double violation of CHSH inequality with
equal sharpness of measurements has also been theoretically predicted
\cite{shareBT2,shareBT3,shareBT4} and experimentally observed
\cite{shareBE1,shareBE2,shareBE3}. Further studies showed that an
unbounded number of CHSH violations can be achieved using weak
measurements with unequal sharpness \cite{shareBT5,shareBT6}. The
sequential sharing of tripartite Bell nonlocality \cite{shareBT7},
EPR steering \cite{shareST1,shareST2,shareST3,shareST4,shareST5},
entanglement \cite{shareET1}, and nonlocal advantage of quantum
coherence \cite{sharenaqc1,sharenaqc2} have also been studied via
weak measurement, showing the advantage of considering this
measurement strategy. While the above works are limited to one-sided
measurements, the sharing of Bell nonlocality \cite{twoside1,twoside2},
EPR steering \cite{twoside3},
two-qubit entanglement \cite{twoside4}, as well as the quantum
advantage in generating random access codes \cite{twoside5}
under two-sided measurements for
which an entangled pair is distributed sequentially to multiple
Alices and Bobs attracts growing interest. The nonlocality sharing
via trilateral sequential measurements \cite{threeside} and network
nonlocality sharing \cite{network1,network2,network3,network4} have
also been studied recently.

The sequential sharing of quantum correlations strongly depends on the
measurement precision. The uncertainty principle sets a fundamental
limit on the precise measurements of two incompatible observables on
a particle \cite{EUR1}, and hence can be related to entanglement sharing.
As for its characterization, apart from the conventional one expressed
in terms of the variance of two observables \cite{var}, the entropy
is another preferred quantity and the entropic uncertainty relations
(EURs) have been widely studied in the past decades \cite{EUR1,EUR2}.
In particular, by introducing a quantum memory entangled with the
particle to be measured, Berta \etal \cite{EUR3} proved an
entanglement-assisted EUR for two orthonormal measurements, which
has been verified experimentally \cite{EUR4,EUR5} and leads to
applications including quantum cryptography \cite{EUR3}, entanglement
witness \cite{EUR3,EUR4,EUR5}, and teleportation \cite{Huf1}. It
also attracted growing interest in studying it from various aspects
\cite{EUR1}, especially its connections with quantum correlations
\cite{Pati,Huf2,Huf3} and quantum coherence \cite{qc1,qc2}. The
entanglement-assisted EUR also holds if one of the positive
operator-valued measures (POVMs) has rank-one elements \cite{Coles}.
For arbitrary POVMs, a generalized EUR was obtained by Frank and
Lieb \cite{CMP}.

In this paper, we address the question as to how many observers
can sequentially and independently sharing the entanglement in a
$(d\times d)$-dimensional (i.e., two-qudit) state. Different from the
two-qubit case \cite{shareET1}, we consider this problem from the
perspective of EUR, and this enables us to explore the two-qudit
states which include the two-qubit states as a special case. Moreover,
it also enables us to explore entanglement sharing in the two-sided
measurements scenario where the entangled qudit pair is distributed
sequentially to multiple Alices and Bobs. Compared to the detection
of entanglement using witness operators which,
in general,
needs to collect statistics of the joint measurements on two
qudits to infer the expectation values of the witness operators,
the advantage of using EUR also lies in that it is sufficient
to estimate the probabilities of different outcomes of the
local measurements on each of the qudits \cite{EUR3}.
Although under specific circumstances, the witness operators can
be decomposed as a sum of local operators for which the outcomes of
local measurements are also sufficient to infer the expectation values
of the witness observables \cite{shareET1,newref}, the EUR is still a
feasible complement to the usual approach for detecting entanglement.
In particular, the efficiency of different approaches may be different,
and the construction and decomposition of a general witness operator
for high-dimensional systems are not very easy tasks \cite{newpra}.
We will consider both the
one- and two-sided weak measurements with different pointers and
show that the number of observers sharing the entanglement strongly
depends on the measurement scenarios, the pointer states of the
apparatus, and the local dimension of each qudit. These results not
only provide an alternative dimension for the investigation of
entanglement sharing, but can also shed light on the interplay
between entanglement and quantum measurements for high-dimensional
systems.

\section{Entanglement-assisted EUR for general POVMs} \label{sec:2}
We consider two POVMs $\mathds{X}$ with elements $\{\mathds{X}_x\}$
and $\mathds{Z}$ with elements $\{\mathds{Z}_z\}$. Defining
$X_x = \sqrt{\mathds{X}_x}$ and $Z_z = \sqrt{\mathds{Z}_z}$ the
measurement operators corresponding to  $\mathds{X}_x$ and
$\mathds{Z}_z$, respectively, then for a two-qudit state $\rho_{AB}$,
the entanglement-assisted EUR reads \cite{CMP}
\begin{equation}\label{eq2-1}
 H(X|B)+H(Z|B) \geqslant \log_2 \frac{1}{c''}+H(A|B),
\end{equation}
where $c^{\prime\prime}= \max_{x,z}\tr\left(\mathds{X}_x \mathds{Z}_z\right)$
quantifies the incompatibility of $\mathds{X}$ and $\mathds{Z}$, the
conditional entropy $H(A|B)= S(\rho_{AB})-S(\rho_B)$ with
$S(\rho_{AB})= -\tr(\rho_{AB}\log_2\rho_{AB})$, and likewise for
$S(\rho_B)$ of the reduced state $\rho_B=\tr_A\rho_{AB}$. Moreover,
$H(X|B)$ and $H(Z|B)$ are given by \cite{CMP}
\begin{equation}\label{eq2-2}
\begin{aligned}
 & H(X|B)= -\sum_{x=0}^{d-1} \tr(\tilde{\rho}_{B|X_x}
            \log_2 \tilde{\rho}_{B|X_x})- S(\rho_B), \\
 & H(Z|B)= -\sum_{z=0}^{d-1} \tr(\tilde{\rho}_{B|Z_z}
            \log_2 \tilde{\rho}_{B|Z_z})- S(\rho_B),
\end{aligned}
\end{equation}
where $\tilde{\rho}_{B|X_x}$ and $\tilde{\rho}_{B|Z_z}$ are the
(unnormalized) postmeasurement states of qudit $B$ given by
\begin{equation}\label{eq2-3}
 \tilde{\rho}_{B|X_x}= \tr_A (X_x \rho_{AB} X_x^\dag),~
 \tilde{\rho}_{B|Z_z}= \tr_A (Z_z \rho_{AB} Z_z^\dag),
\end{equation}
where we omitted the identity operators whenever their presence
is implied by context, e.g., $X_x \rho_{AB} X_x^\dag$ should be
understood as $(X_x\otimes\iden)\rho_{AB}(X_x^\dag\otimes\iden)$.
Note that $H(X|B)$ does not equal the conditional entropy of
$\sum_x X_x \rho_{AB} X_x^\dag$ (i.e., the output of $\mathds{X}$
without postselection) if $\mathds{X}$ is not a rank-one measurement,
and likewise for $H(Z|B)$ \cite{CMP}. In addition, for the rank-one
orthogonal-projective measurements, Eq. \eqref{eq2-1} reduces to the
entanglement-assisted EUR given by Berta \etal \cite{EUR3}.

As $-H(A|B)$ gives the lower bound of the one-way distillable
entanglement in state $\rho_{AB}$ \cite{Cerf,Devetak}, one can
see from Eq. \eqref{eq2-1} that whenever the uncertainty
$U^\mathrm{OS}_{AB} \coloneqq H(X|B)+ H(Z|B)$ estimated by the
one-sided measurement on qudit $A$ and state tomography on qudit $B$
is smaller than $\log_2(1/c'')$, then $\rho_{AB}$ is entangled.
Moreover, as quantum measurements never decrease entropy,
$U^\mathrm{TS}_{AB} \coloneqq H(X|X) + H(Z|Z)$, which corresponds
to the uncertainty estimated by direct two-sided measurements and is
favored for its ease of implementation, provides an upper bound of
$U^\mathrm{OS}_{AB}$. As a consequence, $U^{\mathrm{TS}}_{AB}< \log_2 (1/c'')$
can also be used for entanglement witness. Here, $H(X|X)$ and $H(Z|Z)$
can be obtained as follows \cite{EUR4}:
\begin{equation}\label{eq2-4}
\begin{aligned}
 & H(X|X)= -\sum_{x_1,x_2=0}^{d-1} p_{x_1 x_2}\log_2 p_{x_1 x_2}
           +\sum_{x_2=0}^{d-1} p_{x_2} \log_2 p_{x_2}, \\
 & H(Z|Z)= -\sum_{z_1,z_2=0}^{d-1} q_{z_1 z_2}\log_2 q_{z_1 z_2}
           +\sum_{z_2=0}^{d-1} q_{z_2} \log_2 q_{z_2},
\end{aligned}
\end{equation}
where the probability distribution
$p_{x_1 x_2}= \tr[\rho_{AB}(\mathds{X}_{x_1}\otimes\mathds{X}_{x_2})]$,
$p_{x_2}=\sum_{x_1} p_{x_1 x_2}$, and likewise for $q_{z_1 z_2}$ and
$q_{z_2}$.

\section{Weak measurements on two-qudit states} \label{sec:3}
For a weak measurement, the system is weakly coupled to a pointer
that serves as the measurement apparatus, thus it provides less
information about the system while producing less disturbance
\cite{von,weak1,weak2}. Defining $\{\Pi_i\}$ the projectors in the
basis $\{|i\rangle\}$, $|\phi\rangle$ the initial pointer state, and
$\{|\phi_i\rangle\}$ the evolved pointer states associated with
$\{|i\rangle\}$, then for the initial system state $\rho_0$, the
system-pointer state after the coupling is given by the following map:
\begin{equation}\label{eq3-1}
 \mathcal{E}\left(\rho_0\otimes|\phi\rangle\langle\phi|\right)=
 \sum_{ij} \Pi_i\rho_0\Pi_j \otimes |\phi_i\rangle\langle\phi_j|,
\end{equation}
and by tracing out the pointer one can obtain the nonselective
postmeasurement state as $\rho= \sum_{ij} \Pi_i\rho_0\Pi_j
\langle\phi_i|\phi_j\rangle$.

We consider the two-qudit state $\rho_{AB}$. Defining
$F= \langle\phi_i|\phi_{j\neq i}\rangle$ (it depends on $i$ and $j$ in
general, here for simplicity, supposing it is a constant) the quality
factor measuring the disturbance of a measurement \cite{shareBE1}.
Furthermore, we denote by $\mathds{E}$ the POVM defining the weak
measurement. Then the nonselective postmeasurement states of the
one-sided measurement on $A$ and two-sided measurements on $AB$,
after tracing out the pointer, can be obtained as \cite{sharenaqc2}
\begin{equation}\label{eq3-2}
\begin{aligned}
 \mathds{E}(\rho_{AB})= \, & F\rho_{AB}+(1-F)\sum_i \Pi_i \rho_{AB} \Pi_i, \\
 \mathds{E}^{\otimes 2}(\rho_{AB})= \, & F\rho_{AB}+(1-F) \sum_{i,j}
                                         (\Pi_i\otimes\Pi_j)\rho_{AB}(\Pi_i\otimes\Pi_j) \\
                                       & +(F^2-F) \sum_{i\neq k,j\neq l}
                                         (\Pi_i\otimes\Pi_j)\rho_{AB}(\Pi_k\otimes\Pi_l).
\end{aligned}
\end{equation}

As the pointer states $\{|\phi_i\rangle\}$ are not perfectly
distinguishable, we further choose a complete orthogonal set
$\{|\varphi_i\rangle\}$ as the reading states. Then for the
one-sided case, the probability $p_i$ of getting outcome $i$
and the associated (unnormalized) collapsed state
$\tilde{\rho}_{AB|i}= \langle\varphi_i|\mathcal{E}(\rho_{AB}
\otimes |\phi\rangle\langle\phi|)|\varphi_i\rangle$ are given by
\begin{equation}\label{eq3-3}
\begin{aligned}
 p_i=\, & \tr(\Pi_i\rho_{AB})|\langle\varphi_i|\phi_i\rangle|^2
          +\sum_{j\neq i}\tr(\Pi_j\rho_{AB})
          |\langle\varphi_i|\phi_j\rangle|^2, \\
 \tilde{\rho}_{AB|i}=\, & \Pi_i\rho_{AB}\Pi_i |\langle\varphi_i|\phi_i\rangle|^2
                          +\sum_{j\neq i} \Pi_j\rho_{AB}\Pi_j
                          |\langle\varphi_i|\phi_j\rangle|^2 \\
                        & +\sum_{k\neq l} \Pi_k\rho_{AB}\Pi_l
                          \langle\varphi_i|\phi_k\rangle
                          \langle\phi_l|\varphi_i\rangle,
\end{aligned}
\end{equation}
and without loss of generality, we suppose the measurements are unbiased.
Here, by saying the measurements are unbiased, we mean that the pointer
states $\{|\phi_i\rangle\}$ and reading states $\{|\varphi_i\rangle\}$ yield the same
$|\langle\varphi_i|\phi_i\rangle|^2$ ($\forall i$) and the same
$|\langle\varphi_i|\phi_{j\neq i}\rangle|^2$ ($\forall j\neq i$), that is,
$|\langle\varphi_i|\phi_i\rangle|^2$ is independent of $i$ and
$|\langle\varphi_i|\phi_{j\neq i}\rangle|^2$ is independent of $j\neq i$,
then Eq. \eqref{eq3-3} can be rewritten as
\begin{equation}\label{eq3-4}
\begin{aligned}
 p_i= \, & G\tr(\Pi_i\rho_{AB})+\frac{1-G}{d}, \\
 \tilde{\rho}_{AB|i}= \, & \frac{\mathcal{F}}{d}\rho_{AB}
                           +\frac{1+d_1G- \mathcal{F}}{d}\Pi_i\rho_{AB}\Pi_i \\
                         & +\frac{1-G-\mathcal{F}}{d}\left(\sum_{j\neq i}\Pi_j\rho_{AB}\Pi_j
                           +\sum_{k\neq l\atop k,l\neq i}\Pi_k\rho_{AB}\Pi_l \right),
\end{aligned}
\end{equation}
where we defined $d_1=d-1$, $\mathcal{F}=[(1+d_1G)(1-G)]^{1/2}$,
and $G= 1-d|\langle\varphi_i|\phi_{j\neq i}\rangle|^2$ measures the
precision (or equivalently, the information gain) of the weak
measurement. From Eq. \eqref{eq3-4}, one can obtain
$\tilde{\rho}_{B|i}= \tr_A\tilde{\rho}_{AB|i}$ which will be used to
calculate $H(X|B)$ and $H(Z|B)$.

Similarly, for the two-sided measurements, the probability distribution
$p_{ij}$ of the measurement outcomes $i$ on $A$ and $j$ on $B$ can be
obtained as
\begin{equation}\label{eq3-5}
\begin{aligned}
 p_{ij}= \, & \tr[(\Pi_i\otimes\Pi_j)\rho_{AB}]
              |\langle\varphi_i|\phi_i\rangle|^2 |\langle\varphi_j|\phi_j\rangle|^2 \\
            & +\sum_{k\neq i} \tr[(\Pi_k\otimes\Pi_j)\rho_{AB}]
              |\langle\varphi_i|\phi_{k}\rangle|^2 |\langle\varphi_j|\phi_j\rangle|^2 \\
            & +\sum_{l\neq j} \tr[(\Pi_i\otimes\Pi_l)\rho_{AB}]
              |\langle\varphi_i|\phi_i\rangle|^2 |\langle\varphi_j|\phi_{l}\rangle|^2 \\
            & +\sum_{k\neq i,l\neq j} \tr[(\Pi_k\otimes\Pi_l)\rho_{AB}]
              |\langle\varphi_i|\phi_{k}\rangle|^2 |\langle\varphi_j|\phi_{l}\rangle|^2,
\end{aligned}
\end{equation}
then by defining $\tilde{\Pi}_i=\iden-\Pi_i$ and using $\mathcal{F}$
and $G$ defined above, Eq. \eqref{eq3-5} can be reformulated as
\begin{equation}\label{eq3-6}
\begin{aligned}
 p_{ij}= \, & \frac{(1+d_1G)^2}{d^2} \tr[(\Pi_i\otimes\Pi_j)\rho_{AB}] \\
            & +\frac{(1-G)^2}{d^2} \tr[(\tilde{\Pi}_i\otimes\tilde{\Pi}_j)\rho_{AB}] \\
            & +\frac{\mathcal{F}^2}{d^2} \tr[(\Pi_i\otimes\tilde{\Pi}_j
              +\tilde{\Pi}_i\otimes\Pi_j)\rho_{AB}],
\end{aligned}
\end{equation}
and this will be used to calculate $H(X|X)$ and $H(Z|Z)$.

As we are interested in entanglement sharing, we consider
$\Pi_x= |\psi^d_x\rangle \langle\psi^d_x|$ and
$\Pi_z= |\psi^0_z\rangle \langle\psi^0_z|$ ($x,z=0,1,\ldots,d-1$)
the projectors associated with
$\mathds{X}$ and $\mathds{Z}$ defining two weak measurements,
where $\{|\psi^d_x\rangle\}$ and $\{|\psi^0_z\rangle\}$ are two
sets of mutually unbiased bases (MUBs). For $d=2$, the three MUBs
$\{|\psi^2_x\rangle\}$, $\{|\psi^1_y\rangle\}$, and
$\{|\psi^0_z\rangle\}$ are eigenbases of the Pauli operators
$\sigma_1$, $\sigma_2$, and $\sigma_3$, respectively. For $d>2$,
we construct them via the following MUBs:
\begin{equation}\label{eq3-7}
 \left|\psi^d_x\right\rangle= \frac{1}{\sqrt{d}} \sum_{n=0}^{d-1} e^{i\frac{2\pi}{d}xn} |n\rangle,~
 \left|\psi^0_z\right\rangle= \sum_{n=0}^{d-1} \delta_{zn} |n\rangle,
\end{equation}
where $\delta_{zn}$ and $i$ represent the Delta function and the
imaginary unit, respectively. Moreover, for any prime $d\geqslant 3$,
one can also construct $\mathds{X}$ and $\mathds{Z}$ via any pair of
the MUBs. In addition to $\{|\psi^d_x\rangle\}$ and $\{|\psi^0_z\rangle\}$
in Eq. \eqref{eq3-7}, the remaining $d-1$ MUBs $\{|\psi^r_a\rangle\}$
($r=1,\ldots,d-1$) are given by \cite{MUB1,MUB2}:
\begin{equation}\label{eq3-8}
 \left|\psi^r_a\right\rangle= \frac{1}{\sqrt{d}} \sum_{n=0}^{d-1}
                              e^{i\frac{2\pi}{d}r(a+n)^2} |n\rangle ~
                              (r=1, \ldots, d-1),
\end{equation}
where the index $a\in \{0,1,\ldots,d-1\}$ for any given $r$.

For the weak measurements, there is a trade-off $F^2+G^2\leqslant1$
for two-qubit systems \cite{shareBT1}. For a two-qudit system, as $F$
and $G$ defined above are basis independent, one can write the reading
and pointer states as $|\varphi_i\rangle=|i\rangle$ and
$|\phi_i\rangle=u_{i}|i\rangle+ \sum_{k \neq i}v_{i k}|k\rangle$, respectively.
By combining this with the assumption that the measurements are unbiased, one has
$|u_{i}|\equiv u$ ($\forall i$), $|v_{i k}|\equiv v$ ($\forall k\neq i$), and
$u^2+d_1 v^2=1$, which gives $F= |u_{i}v_{j i}+u_{j}v_{i j}+\sum_{k\neq i,j}v_{i k}v_{j k}|$
and $G=1-dv^2$. Hence
\begin{equation}\label{eq3-new1}
 F^2+G^2 \leqslant (2uv+d_2 v^2)^2+(1-dv^2)^2 \leqslant 1,
\end{equation}
that is, one still has the trade-off $F^2+G^2\leqslant1$.
When the equality holds, the pointer is said to be optimal
in the sense that the highest precision is achievable for
a given quality factor, and the corresponding optimal
pointer can be obtained by solving the equation
$(2uv+d_2 v^2)^2+(1-dv^2)^2 = 1$. But for the general weak
measurements (e.g., the biased case), whether or not the trade-off
$F^2+G^2\leqslant1$ holds still remains an open question.

In addition, there are other pointers, including the square and
Gaussian pointers \cite{shareBT1}. The unsharp measurement is
another kind of weak measurement \cite{unsharp}, and when it
does not lead to confusion we also term it as the weak measurement
with an unsharp pointer. The associated POVM elements of
$\mathds{X}$ and $\mathds{Z}$ are given by \cite{shareST2}
\begin{equation}\label{eq3-9}
 \mathds{X}_x= \lambda\Pi_x+ \frac{1-\lambda}{d}\iden,~
 \mathds{Z}_z= \lambda\Pi_z+ \frac{1-\lambda}{d}\iden,
\end{equation}
where $ \lambda\in[0,1]$ is the sharpness parameter. For $\mathds{X}$
constructed by the MUB $\{\Pi_x\}$, one has \cite{shareST2}
\begin{equation}\label{eq3-10}
 X_x= \left(\sqrt{\frac{1+d_1\lambda}{d}}
      -\sqrt{\frac{1-\lambda}{d}}\right)\Pi_x
      +\sqrt{\frac{1-\lambda}{d}}\iden,
\end{equation}
and likewise for $Z_z$. In the language of weak measurements, the
reading and pointer states of the unsharp measurement can be written as
 $|\varphi_i\rangle=|i\rangle$
and $|\phi_i\rangle=\sqrt{1-d_1 u^2}|i\rangle+ u\sum_{j\neq i}|j\rangle$ ($\forall i$), respectively, where
$u= \sqrt{(1-\lambda)/d}$ \cite{twoside3}. In addition, by defining
$d_2=d-2$, the quality factor $F$ and measurement precision $G$ of the unsharp
measurement can be obtained as
\begin{equation}\label{eq3-11}
  F= \frac{d_2(1-\lambda)+ 2\sqrt{1+d_2\lambda-d_1\lambda^2}}{d},~
  G= \lambda,
\end{equation}
thus it is optimal for $d=2$ and nonoptimal for $d\geqslant 3$.

Supposing that $\{\mathds{E}_i\}$ are the elements of the POVM
$\mathds{E}$ defining the weak measurement, then one also has
$p_i=\tr(\mathds{E}_i\rho_{AB})$. Combining this with Eq. \eqref{eq3-4}
and taking $\rho_{AB}=\iden\otimes\iden/d^2$ yields $\tr \mathds{E}_i=1$
($\forall i$). Then for $\mathds{X}$ and $\mathds{Z}$ with the elements
$\mathds{X}_x=r_0\iden/d+ \sum_{m} r_{x m} |\psi^r_m\rangle\langle\psi^r_m|$ and
$\mathds{Z}_z= s_0\iden/d+ \sum_{n}s_{z n} |\psi^s_n\rangle\langle\psi^s_n|$
comprising of the identity operator $\iden$ and the MUBs
$\{|\psi^r_m\rangle\}$ and $\{|\psi^s_n\rangle\}$, one can obtain
$\tr(\mathds{X}_x\mathds{Z}_z)= 1/d$ ($\forall x,z$), thus
$c^{\prime\prime}= 1/d$. Here, the coefficients $r_0$ and $r_{x m}$
satisfy $\sum_x \mathds{X}_x=\iden$ and $\tr\mathds{X}_x=1$ ($\forall x$),
and likewise for $s_0$ and $s_{z n}$.

In the following, we mainly concentrate on $\mathds{X}$ and
$\mathds{Z}$ constructed by $\{|\psi^d_x\rangle\}$ and
$\{|\psi^0_z\rangle\}$, respectively, as we aim at exploring the
two-qudit state with $d$ being a general positive integer. When $d$
is a prime, we will also sketch the main results for the cases that
$\mathds{X}$ and $\mathds{Z}$ are constructed via other MUBs at the
end of this paper.

\section{Entanglement sharing of two-qudit states} \label{sec:4}
\begin{figure}
\centering
\resizebox{0.47 \textwidth}{!}{%
\includegraphics{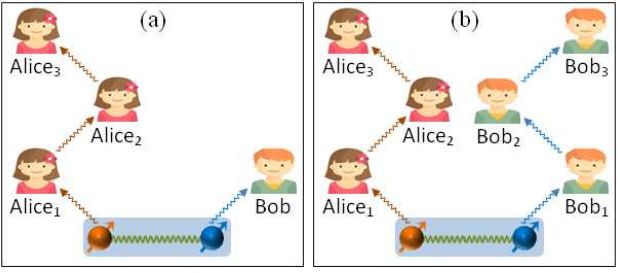}}
\caption{Two scenarios of entanglement sharing via EUR. (a) Multiple
Alices measure sequentially on their side  and Bob checks whether or not
there is shared entanglement via the uncertainties of Alices' outcomes
estimated by his measurements. (b) Multiple Alices and Bobs measure
sequentially on their respective sides, after which they check whether
there is shared entanglement via the uncertainties estimated by their
measurement statistics.} \label{fig:1}
\end{figure}

In this work, we consider entanglement sharing via both the one- and
two-sided sequential weak measurements. As shown in Fig. \ref{fig:1}(a),
the one-sided scenario refers to the case in which multiple Alices
(say, Alice$_1$, Alice$_2$, etc.) have access to half of an entangled
pair and a single Bob has access to the other half. Alice$_1$ performs
her randomly selected measurement ($\mathds{X}$ or $\mathds{Z}$) and
records the outcome. She then passes her qudit to Alice$_2$ who also
measures randomly $\mathds{X}$ or $\mathds{Z}$ on the received qudit,
records the outcome, and passes it to Alice$_3$, and so on. For the
two-sided scenario, as shown in Fig. \ref{fig:1}(b), multiple Alices
(Alice$_1$, Alice$_2$, etc.) have access to half of an entangled pair
and multiple Bobs (Bob$_1$, Bob$_2$, etc.) have access to the other
half. To proceed, Alice$_1$ and Bob$_1$ choose randomly the same
POVM $\mathds{X}$ or $\mathds{Z}$, perform measurements and record
their outcomes. They then pass their qudits to Alice$_2$ and Bob$_2$
who repeat this process again, and so on.

For each measurement scenario, we further consider two slightly
different cases for which we illustrate through the one-sided scenario
(they are similar for the two-sided scenario). In the first case, the
classical information pertaining to each Alice's measurement choice
and outcome is not shared, thus the state shared between Alice$_n$
($n\geqslant 2$) and Bob is averaged over all the possible inputs and
outputs of Alice$_{n-1}$, from which one arrives at Bob's uncertainty
about Alice$_n$'s outcome. In the second case, however, each Alice
informs the next Alice of her measurement choice but not the outcome,
thus Bob's uncertainty has to be averaged over the uncertainty for
each of their possible inputs. More specifically, for the first case
one first obtain the average state of all the possible inputs and
then arrives at the associated uncertainty, while for the second
case one first obtain the uncertainty associated with each possible
input and then arrives at their average effect. For convenience of
later presentation, we term these two cases as scenario OS1 and
scenario OS2, respectively. Similarly, the two cases associated
with the two-sided scenario are termed as scenario TS1 and scenario
TS2, respectively.

We would like to mention that when exploring nonlocality, EPR steering,
and entanglement, both the scenarios OS1 \cite{shareBT1,shareBT2,
shareBT3,shareBT4,shareBT5,shareST2,shareST3,shareET1} and TS1
\cite{twoside1,twoside2,twoside3,twoside4} have been
considered previously.
Moreover, scenario OS2 is analogous to a scenario of
entanglement-assisted EUR which is useful in cryptographic protocols
such as quantum key distribution
\cite{EUR1,EUR3}. Due to the fact that the multiple Alices (Bobs)
in each side can be spatially separated, these scenarios are
expected to play a similar role in multipartite cryptographic tasks.

Having collected all the required tools for our
analysis, we are now in a position to investigate sequential
sharing of entanglement for the aforementioned scenarios.
To set the stage, we start by indicating some notations that
will be employed in the following discussion. We
denote by $\mathds{E}^{(0)}= \iden$ the identity map and
$\mathds{E}^{(n)} \in \{\mathds{X}, \mathds{Z}\}$ ($n=1,2,\ldots$)
the POVMs defining the measurements of Alice$_n$ and Bob$_n$.
Moreover, we denote by $F_n$ ($G_n$) the quality factor
(precision) of the corresponding weak measurement and
$G_{n,c}^{\mathrm{OS}}$ ($G_{n,c}^{\mathrm{TS}}$)
the critical value of $G_n$ larger than which the entanglement
could be shared by Alice$_n$ and Bob (Bob$_n$),
provided that the previous Alices and Bob (Bobs) have already shared
the entanglement. Finally, we denote by
$|\Psi^{+}\rangle= \sum_{k=0}^{d-1}|kk\rangle/\sqrt{d}$ the two-qudit
maximally entangled state shared between Alice$_1$ and Bob
(one-sided case) or Alice$_1$ and Bob$_1$ (two-sided case).

\subsection{One-sided scenario} \label{sec:4a}
We first consider the scenario OS1. If the initial state shared
between Alice$_1$ and Bob is $\rho_{A_1B}$, then the state shared
between Alice$_n$ and Bob will be given by
\begin{equation}\label{eq4a-1}
 \rho_{A_n B}= \frac{1}{2^{n-1}} \sum_{\mathds{E}^{(0)}, \ldots, \mathds{E}^{(n-1)}}
               \mathds{E}^{(n-1)} \circ \ldots \circ \mathds{E}^{(0)} (\rho_{A_1B}),
\end{equation}
where the factor $1/2^{n-1}$ is due to the fact that $\rho_{A_n B}$
is an average state of the $2^{n-1}$ possible inputs of Alice$_n$,
while $\mathds{E}^{(0)}$ is introduced for ensuring that $\rho_{A_n B}$
reduces to $\rho_{A_1 B}$ when $n = 1$.
For the general weak measurement, ${E}^{(1)} (\rho_{A_1B})$ can
be obtained from Eq. \eqref{eq3-2}, and likewise for
$\mathds{E}^{(n-1)} \circ \ldots \circ \mathds{E}^{(0)} (\rho_{A_1B})$.
As the POVMs $\mathds{E}^{(l)} \in \{\mathds{X}, \mathds{Z}\}$ ($l=1,\ldots,n-1$),
when the elements $\{\mathds{X}_x\}$ of $\mathds{X}$ and $\{\mathds{Z}_z\}$ of $\mathds{Z}$
are known, by defining $X_x= \sqrt{\mathds{X}_x}$ and $Z_z= \sqrt{\mathds{Z}_z}$,
one also has
\begin{equation}\label{eq4-new1}
 \rho_{A_2B}= \frac{1}{2}\left(\sum_x X_x \rho_{A_1B} X_x^\dag
             +\sum_z Z_z \rho_{A_1B} Z_z^\dag \right),
\end{equation}
and likewise for $\rho_{A_n B}$ \cite{Luders}. Moreover,
$\rho_{A_n B}$ in Eq. \eqref{eq4a-1} is normalized. This is because
$\mathds{E}^{(n-1)} \circ \ldots \circ \mathds{E}^{(0)} (\rho_{A_1B})$
is a nonselective postmeasurement state which is normalized and different from
the selective postmeasurement state, e.g., for the POVM $\mathds{X}$ and initial
state $\rho_{A_1B}$, the postmeasurement state associated with the outcome $x$ is
$X_x \rho_{A_1B} X_x^\dag/\tr(\mathds{X}_x \rho_{A_1B})$ \cite{Luders}.

In the following, we choose
$\rho_{A_1 B}= |\Psi^{+}\rangle\langle \Psi^{+}|$.
Then from Eqs. \eqref{eq4a-1} and \eqref{eq3-4} one can obtain
$\tilde{\rho}_{A_nB|\mathds{X}_x}$ and $\tilde{\rho}_{A_nB|\mathds{Z}_z}$.
A direct calculation shows that for both $\tilde{\rho}_{B|\mathds{X}_x}=
\tr_{A_n}\tilde{\rho}_{A_nB|\mathds{X}_x}$ ($\forall x$) and
$\tilde{\rho}_{B|\mathds{Z}_z}= \tr_{A_n} \tilde{\rho}_{A_nB|\mathds{Z}_z}$
($\forall z$), the eigenvalues are given by
\begin{equation}\label{eq4a-2}
 \epsilon_0= \frac{1+d_1 \mu_n}{d^2}~(1),~
 \epsilon_1= \frac{1-\mu_n}{d^2} ~(d_1),
\end{equation}
where the numbers in the brackets denote the degeneracy, and
by defining $F_0=1$, the parameter $\mu_n$ can be written as
\begin{equation}\label{eq4a-3}
 \mu_n=\frac{G_n\prod_{k=0}^{n-1}(1+F_k)}{2^{n}}.
\end{equation}
As $\rho_B= \iden/d$, from Eqs. \eqref{eq2-2} and \eqref{eq4a-2},
one arrives at the uncertainty
\begin{equation}\label{eq4a-4}
 U^{\mathrm{OS}}_{A_nB}= 2\left[H_2\left(\frac{1+d_1 \mu_n}{d}\right)
                         +\frac{d_1(1-\mu_n)}{d}\log_2 d_1\right],
\end{equation}
where $H_2(\cdot)$ denotes the binary Shannon entropy function.

\begin{figure}
\centering
\resizebox{0.44 \textwidth}{!}{%
\includegraphics{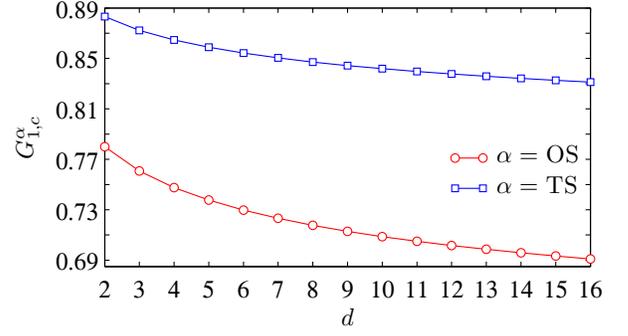}}
\caption{The critical precision $G_{1,c}^\mathrm{\alpha}$
($\alpha= \mathrm{OS}$ or $\mathrm{TS}$) versus $d$ for the weak
measurement.} \label{fig:2}
\end{figure}

Based on Eq. \eqref{eq4a-4}, one can determine the maximum number of
Alices  sharing the entanglement with Bob by checking whether or not
$U^{\mathrm{OS}}_{A_nB}< \log_2 d$ holds. As $U^{\mathrm{OS}}_{A_nB}$
is a monotonic decreasing function of $G_n$,
the measurement precision
must be larger than a critical value $G_{n,c}^{\mathrm{OS}}$ for
achieving possible entanglement sharing.
Of course, $G_{n,c}^{\mathrm{OS}}$ with $n\geqslant 2$
is not a constant as the actual precision $G_l$ of Alice$_l$
($l\leqslant n-1$) may be larger than $G_{l,c}^{\mathrm{OS}}$.
As for $G_{1,c}^{\mathrm{OS}}$, its value could be obtained by
solving numerically the transcendental equation
$U^{\mathrm{OS}}_{A_1B}= \log_2 d$, and it is obvious that it is
independent of $F_1$.
This is because $U^{\mathrm{OS}}_{A_1B}$ depends on the initial state
$\rho_{A_1B}$ and the measurement precision $G_1$  (i.e., the information
gain) of Alice$_1$, while the disturbance of her measurement affects only the
output state of the qudit which is passed on to Alice$_2$, and this will
further affects $U^{\mathrm{OS}}_{A_2B}$ and $G_{2,c}^{\mathrm{OS}}$.
As illustrated in Fig. \ref{fig:2}, $G_{1,c}^{\mathrm{OS}}$ decreases with
the increase of $d$, and for infinite large $d$, it approaches the asymptotic
value $G_{1,c}^{\mathrm{OS}}(\infty)= 1/2$. This shows that the larger the
local dimension $d$ of each qudit, the lower the critical measurement
precision is required for sharing the entanglement.

When the measured qudit is passed to the subsequent Alices, the
uncertainties $U^{\mathrm{OS}}_{A_nB}$ will be dependent on the quality
factors $F_l$ ($l=1,\ldots,n-1$) of the measurements. Based on Eq. \eqref{eq4a-4}, one can see
that the maximum number of Alices sharing the entanglement with Bob
can be determined by checking whether the following inequalities hold 
\begin{equation}\label{eq4a-5}
 G_l\geqslant G_{l,c}^\mathrm{OS}~(l=1,\ldots,n-1),~
 G_{n,c}^\mathrm{OS}= \frac{2^{n}G_{1,c}^\mathrm{OS}}{\prod_{k=0}^{n-1}(1+F_k)}\leqslant 1,
\end{equation}
where the first inequality ensures that Alice$_l$ ($l=1,\ldots,n-1$)
share the entanglement simultaneously. Moreover, as $U^{\mathrm{OS}}_{A_nB}$
is a decreasing function of $\mu_n$ and $\mu_n$ of Eq. \eqref{eq4a-3}
is an increasing function of $G_n$, one must have
$\mu_{n,c}^\mathrm{OS}= \mu_{1,c}^\mathrm{OS}$ ($\mu_{n,c}^\mathrm{OS}$
is the critical $\mu_n$ obtained by replacing $G_n$ with
$G_{n,c}^\mathrm{OS}$) to ensure that Alice$_n$ can also share the
entanglement. This, together with Eq. \eqref{eq4a-3}, gives the above
equality. As for the second inequality, it is due to the fact that
$G_n\leqslant 1$ ($\forall n$) by its definition.

We now discuss whether or not the inequalities in Eq. \eqref{eq4a-5}
hold when the qudit is passed to Alice$_2$. We consider the weak
measurements with unsharp, optimal, Gaussian, and square pointers.

(A1.1) For the unsharp pointer, the relation between $F_1$ and $G_1$
is in Eq. \eqref{eq3-11}, by combining of which with Eq. \eqref{eq4a-5}
one can obtain that when $G_{1,c}^\mathrm{OS}\leqslant G_1\leqslant 2\xi/d$,
there exists valid $G_{2,c}^\mathrm{OS}$ for all $d\geqslant 2$,
where the parameter $\xi$ is given by
\begin{equation}\label{eq4a-6}
 \xi= d_2\left(1-G_{1,c}^\mathrm{OS}\right)
      +\sqrt{2\left(1-G_{1,c}^\mathrm{OS}\right)
      \left(2d_1G_{1,c}^\mathrm{OS}-d_2\right)}.
\end{equation}

(A1.2) For the optimal pointer, $F_1^2+G_1^2=1$, then from Eq.
\eqref{eq4a-5} one arrives at $G_{1,c}^\mathrm{OS} \leqslant
G_1 \leqslant 2[G_{1,c}^\mathrm{OS}(1-G_{1,c}^\mathrm{OS})]^{1/2}$,
which holds if $G_{1,c}^\mathrm{OS} \leqslant 0.8$. As
$G_{1,c}^\mathrm{OS}\simeq 0.7799$ for $d=2$ and decreases with the
increasing $d$, there exists valid $G_{2,c}^\mathrm{OS}$ for all
$d\geqslant 2$.

(A1.3) For the Gaussian pointer, the relation of $F_1$ and $G_1$
can be obtained numerically \cite{shareBT1}. We denote it by
$\Lambda(F_1)=G_1$, i.e., for any given $F_1$, the map $\Lambda(F_1)$
gives the associated $G_1$. Then from Eq. \eqref{eq4a-5} one can obtain
$G_{1,c}^\mathrm{OS}\leqslant G_1 \leqslant\Lambda(2G_{1,c}^\mathrm{OS}-1)$,
which holds when $G_{1,c}^\mathrm{OS}\lesssim 0.7547$. By combining
this with $G_{1,c}^\mathrm{OS}$ obtained for different local dimension
$d$ (see Fig. \ref{fig:2}), one can see that there exists valid
$G_{2,c}^\mathrm{OS}$ for $d\geqslant 4$.

(A1.4) For the square pointer which is far from optimal, one has
$F_1+G_1=1$ \cite{shareBT1}, then from Eq. \eqref{eq4a-5} one arrives
at $G_{1,c}^\mathrm{OS}\leqslant G_1\leqslant 2-2G_{1,c}^\mathrm{OS}$.
This inequality holds  when $G_{1,c}^\mathrm{OS}\leqslant 2/3$. By
combining this with $G_{1,c}^\mathrm{OS}$ obtained for different $d$,
one can see that there exists valid $G_{2,c}^\mathrm{OS}$ for
$d\geqslant 34$.

The critical $G_{2,c}^\mathrm{OS}$ also decreases with the increase
of $d$ and approaches its asymptotic value
$1/[1+F_{1,c}^{\mathrm{OS}}(\infty)]$ when $d\rightarrow \infty$,
where $F_{1,c}^{\mathrm{OS}}(\infty)$ is the quality factor associated
with $G_{1,c}^{\mathrm{OS}}(\infty)$. Following the same line of
reasoning as above, one can obtain that for the optimal pointer with
$d\geqslant 34$ and the Gaussian pointer with $d\geqslant 141$,
the entanglement could be sequentially shared by at most three Alices
and Bob, and for large enough $d$, the maximum number of Alices could
be further enhanced.  In fact, a similar phenomenon for sequential
sharing of EPR steering has also been observed previously \cite{shareST2}.
But for the unsharp and square pointers, the maximum number of
Alices remains 2. We explain this for the square pointer (the case
is similar for the unsharp pointer). From Eq. \eqref{eq4a-5} one can
obtain that if there exists a valid $G_{3,c}^\mathrm{OS}$, then
$(2-G_1)(2-G_2)\geqslant 4G_{1,c}^\mathrm{OS}$. However, even under
$G_1=G_{1,c}^\mathrm{OS}$ and $G_2=G_{2,c}^\mathrm{OS}$, this
inequality holds only when $G_{1,c}^\mathrm{OS}\leqslant 1/2$. As
illustrated above, $G_{1,c}^\mathrm{OS}\geqslant 1/2$, and the
equality holds when $d\rightarrow \infty$, thus the entanglement
cannot be sequentially shared by three Alices and Bob.

Next, we turn to consider the scenario OS2 for which each Alice knows
the measurement choice but not the outcome of the former Alices, and
our aim is to determine whether or not the entanglement can be shared
by multiple Alices and Bob. We will consider the average effect, that
is, the shared entanglement between Alice$_n$ and Bob is averaged
over the possible inputs and outputs of Alice$_{n-1}$, and likewise for
Bob's uncertainty about Alice$_n$'s outcome. Clearly, this is different
from the scenario OS1 for which the entanglement and uncertainty are
obtained for the state averaged over the previous Alice's possible
inputs and outputs. For this case, the possible states shared
between Alice$_n$ and Bob are given by
\begin{equation}\label{eq4a-7}
 \rho_{A_n B}^{\mathds{E}^{(0)} \ldots \mathds{E}^{(n-1)}}=
                                \mathds{E}^{(n-1)} \circ \ldots \circ
                                \mathds{E}^{(0)}(\rho_{A_1B}),
\end{equation}
where the meanings of $\mathds{E}^{(0)}$ and $\mathds{E}^{(n)}$ ($n\geqslant 1$)
are the same as that in Eq. \eqref{eq4a-1}. For each possible input, from Eq.
\eqref{eq3-4} one can obtain the unnormalized postmeasurement states
$\tilde{\rho}_{A_nB|\mathds{O}_o}^{\mathds{E}^{(0)} \ldots \mathds{E}^{(n-1)}}$,
where $\mathds{O}_o=\mathds{X}_x$ or $\mathds{Z}_z$. Then after a straightforward
but somewhat complex calculation, one can obtain the eigenvalues of
$\tilde{\rho}^{\mathds{E}^{(0)} \ldots \mathds{E}^{(n-1)}}_{B|\mathds{O}_o}=
\tr_{A_n} \tilde{\rho}_{A_nB|\mathds{O}_o}^{\mathds{E}^{(0)} \ldots \mathds{E}^{(n-1)}}$
($\forall o$) as
\begin{equation}\label{eq4a-8}
\begin{aligned}
 & \epsilon_0= \frac{1+d_1G_n \prod_{k=0}^{n-1}F_k^{\delta(\mathds{E}^{(k)},\mathds{O})}}{d^2}~(1),\\
 & \epsilon_1= \frac{1-G_n \prod_{k=0}^{n-1}F_k^{\delta(\mathds{E}^{(k)},\mathds{O})}}{d^2} ~(d_1),
\end{aligned}
\end{equation}
where we define the function $\delta(\mathds{E}^{(k)},\mathds{O})=0$
if $\mathds{E}^{(k)}=\mathds{O}$ and $\delta(\mathds{E}^{(k)},\mathds{O})=1$
if $\mathds{E}^{(k)}\neq \mathds{O}$, and the numbers in the brackets
still denote the degeneracy. Combining this with Eq. \eqref{eq2-2} gives
the average uncertainty
\begin{equation}\label{eq4a-9}
\begin{aligned}
 \mathcal{U}^{\mathrm{OS}}_{A_n B}= \, & \frac{1}{2^{n-1}} \sum_{\mathds{E}^{(0)}, \ldots, \mathds{E}^{(n-1)}}
                                         U^{\mathrm{OS}}_{A_n B}\left(\rho_{A_n B}^{\mathds{E}^{(0)}
                                         \ldots \mathds{E}^{(n-1)}}\right) \\
                                  = \, & \frac{1}{2^{n-2}} \left\{\sum_{i_0, \ldots, i_{n-1}}
                                         H_2\left(\frac{1+d_1 G_n \prod_{k=0}^{n-1} F^{i_k}_k}{d}\right) \right. \\
                                       & \left. +\frac{d_1}{d}\Bigg[2^{n-1}
                                         -\frac{G_n}{2} \prod_{k=0}^{n-1} (1+F_k)\Bigg]\log_2 d_1\right\},
\end{aligned}
\end{equation}
where $i_0=0$ and $i_{k|k\geqslant 1}\in\{0,1\}$ are the power exponents.

\begin{figure}
\centering
\resizebox{0.44 \textwidth}{!}{%
\includegraphics{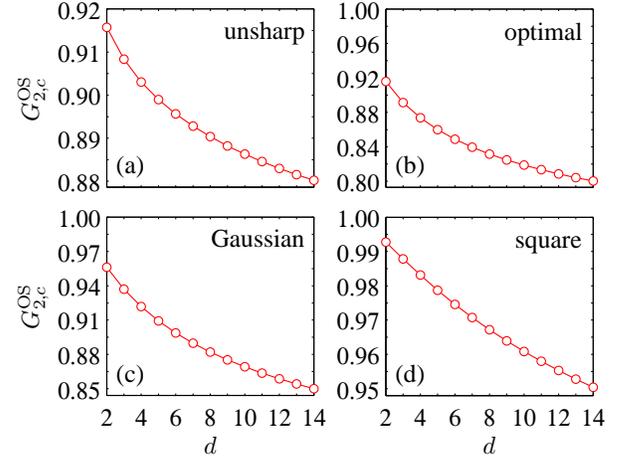}}
\caption{The $d$ dependence of the critical precision $G_{2,c}^\mathrm{OS}$
(obtained with $G_1=G_{1,c}^\mathrm{OS}$) for the scenario OS2 with
different pointers.} \label{fig:3}
\end{figure}

Based on Eq. \eqref{eq4a-9}, one can determine the number of Alices
sharing the average entanglement with one Bob. We focus on $n\geqslant 2$
as the scenario OS2 differentiates the scenario OS1 only for this case.
From Eqs. \eqref{eq4a-1} and \eqref{eq4a-7} one can see that
Alice$_n$'s input for the scenario OS1 can be recognized as an equal
mixture of the possible inputs for the scenario OS2, thus the entanglement
shared between Alice$_n$ and Bob for the scenario OS1 will be weaker
than or equal to that for the scenario OS2 due to the convexity of
entanglement \cite{QE}. As the measurement uncertainty strongly depends
on the amount of entanglement \cite{EUR3}, then intuitively, the
maximum number of Alices sharing the entanglement for the scenario OS2
will be no less than that for the scenario OS1.
In the following, we derive $G_{n,c}^{\mathrm{OS}}$ with
$G_l= G_{l,c}^{\mathrm{OS}}$ ($l=1,\ldots,n-1$), for which the corresponding
quality factor $F_l$ can be obtained for the measurements with
different pointers. By substituting $F_l$ into Eq. \eqref{eq4a-9}
and then solving numerically the transcendental equation
$\mathcal{U}^{\mathrm{OS}}_{A_n B}= \log_2 d$, one can obtain the critical $G_{n,c}^\mathrm{OS}$.
For $n=2$, the corresponding results are
shown in Fig. \ref{fig:3}.

(A2.1) For the unsharp pointer, $G_{2,c}^{\mathrm{OS}}$ approaches the
asymptotic value of $2/3$ when $d\rightarrow \infty$. Moreover, from Eq.
\eqref{eq4a-9} one can check that when the first two Alices share the
average entanglement with Bob, $\mathcal{U}^{\mathrm{OS}}_{A_3B}$ is
always larger than $\log_2 d$. Thus Alice$_3$ cannot share the average
entanglement with Bob.

(A2.2) For the optimal pointer, $G^\mathrm{OS}_{2,c}$ decreases to
$4-2\sqrt{3}$ when $d\rightarrow \infty$. For small $d$, the number
of Alices sharing the entanglement with Bob remains 2. But with the
increasing $d$, the number of Alices may be enhanced. We performed
calculations based on Eq. \eqref{eq4a-9} and it is found that for
$d\geqslant 10$, $180$, and $30608$, respectively, at most three,
four, and five Alices can share the average entanglement with Bob,
and the critical $G_{n,c}^\mathrm{OS}$ ($n=3,4,5$) all decrease with
the increase of $d$. For large enough $d$, it is reasonable to
conjecture that the maximum number of Alices will be further enhanced.

(A2.3) For the Gaussian pointer, a comparison of Fig. \ref{fig:3}(a)
and \ref{fig:3}(c) shows that for $d\geqslant 7$, $G_{2,c}^\mathrm{OS}$ becomes
smaller than that for the unsharp pointer, that is, from the point of
view of entanglement sharing, the Gaussian pointer performs better
than the unsharp pointer for $d\geqslant 7$. A further calculation
shows that for $d\geqslant 67$, the average entanglement can also be
shared by Alice$_3$ and Bob, and the number of Alices can be further
enhanced by increasing $d$. However, the critical $d$ starting from which
the average entanglement can be sequentially shared will be far
larger than that with an optimal pointer.

(A2.4) For the square pointer, $G_{2,c}^\mathrm{OS}$ decreases to
the same asymptotic value as that of the unsharp pointer, which can
be explained from Eq. \eqref{eq3-11} as for the unsharp pointer,
$F_1+G_1 \rightarrow 1$ when $d\rightarrow \infty$. In addition, as its
precision is worse than that of the unsharp pointer, the maximum
number of Alices sharing the average entanglement with Bob remains two.

\subsection{Two-sided scenario} \label{sec:4b}
In this subsection, we discuss sharing of entanglement via
two-sided measurements,
and for simplicity, we focus on the case that both $F_n$ and
$G_n$ ($\forall n$) for Alice$_n$ equal to that for Bob$_n$.
We begin with the scenario TS1 and suppose the state
$\rho_{A_1B_1}= |\Psi^+\rangle\langle\Psi^+|$ is initially shared
between Alice$_1$ and Bob$_1$, then the state $\rho_{A_nB_n}$ shared
between Alice$_n$ and Bob$_n$ takes the
similar form as that given in Eq. \eqref{eq4a-1}, with only
$\mathds{E}^{(k)}$ ($k=0,1, \ldots, n-1$) being replaced by
$\mathds{E}^{(k)}\otimes\mathds{E}^{(k)}$. By combining this with
Eq. \eqref{eq3-6} one can obtain the probability distributions
$p_{x_1 x_2}$ and $q_{z_1 z_2}$ as
\begin{equation}\label{eq4b-1}
\begin{aligned}
 & p_{x_1 x_2|x\in S}= \frac{1+d_1 \nu_n}{d^2},~
   p_{x_1 x_2|x \notin S}= \frac{1-\nu_n}{d^2}, \\
 & q_{z_1 z_1}= \frac{1+d_1 \nu_n}{d^2},~
   q_{z_1 z_2|z_1\neq z_2}= \frac{1- \nu_n}{d^2},
\end{aligned}
\end{equation}
where we define $x=(x_1,x_2)$ and $S=\{(0,0),(n,d-n)|
n= 1, \ldots, d-1 \}$ for simplifying the equations, and $\nu_n$
can be obtained by replacing $G_n$ and $F_k$ in Eq. \eqref{eq4a-3}
with $G_n^2$ and $F_k^2$, respectively.

From Eqs. \eqref{eq4b-1} and \eqref{eq2-4} one can show that the
uncertainty $U^{\mathrm{TS}}_{A_nB_n}$ has a similar form to
$U^{\mathrm{OS}}_{A_nB}$ in Eq. \eqref{eq4a-4}, with only the
parameter $\mu_n$ being replaced by $\nu_n$. So the critical
$G^{\mathrm{TS}}_{1,c}$ larger than which the entanglement can
be shared by Alice$_1$ and Bob$_1$ equals the square root of
$G^{\mathrm{OS}}_{1,c}$, i.e., $G^{\mathrm{TS}}_{1,c}=
(G^{\mathrm{OS}}_{1,c})^{1/2}$, implying a similar behavior for
their dependence on $d$, see Fig. \ref{fig:2}.

Similar to the one-sided scenario, the maximum number of Alices and
Bobs sharing the entanglement can be obtained by checking whether
the following inequalities hold:
\begin{equation}\label{eq4b-2}
\begin{aligned}
  & G_l\geqslant G_{l,c}^\mathrm{TS}~(l=1, \ldots, n-1),\\
  & G_{n,c}^\mathrm{TS}= \frac{2^{n/2}G_{1,c}^\mathrm{TS}} {\sqrt{\prod_{k=0}^{n-1}(1+F_k^2)}} \leqslant 1.
\end{aligned}
\end{equation}

When the two qudits are passed to Alice$_2$ and Bob$_2$,
Eq. \eqref{eq4b-2} reduces to $G_1^2\geqslant G_{1,c}^\mathrm{OS}$
and $F_1^2\geqslant 2G_{1,c}^\mathrm{OS}-1$. We now analyze whether
or not the two inequalities hold simultaneously for the weak
measurements with different pointers.

(B1.1) For the unsharp pointer, the above two inequalities depend on
$d$ and it is hard to give an analytical analysis. The numerical
calculation
by using the trade-off between the quality factor $F_1$
and measurement precision $G_1$ given in Eq. \eqref{eq3-11}
shows that they hold simultaneously when
$G_{1,c}^\mathrm{OS} \lesssim 0.7321$ and $d\gtrsim 224526395$. This
condition is difficult to meet in experiments, hence the unsharp
measurement is not a good choice for achieving entanglement sharing.

(B1.2) For the optimal pointer, the above inequalities are equivalent to
$G_{1,c}^\mathrm{OS} \leqslant G_1^2 \leqslant 2-2G_{1,c}^\mathrm{OS}$,
which holds when $G_{1,c}^\mathrm{OS} \leqslant 2/3$. By combining this
with the results of $G_{1,c}^\mathrm{OS}$ for different $d$, one can
note that the entanglement can be sequentially shared by two Alices and
Bobs for $d\geqslant 34$.

(B1.3) For the Gaussian pointer, based on $\Lambda(F_1)=G_1$, one
can obtain $(G_{1,c}^\mathrm{OS})^{1/2} \leqslant G_1 \leqslant
\Lambda[(2G_{1,c}^\mathrm{OS}-1)^{1/2}]$, which holds for
$G_{1,c}^\mathrm{OS}\lesssim 0.6102$, and this corresponds to
$d\geqslant 404$.

(B1.4) For the square pointer, the above inequalities hold simultaneously
when $G_{1,c}^\mathrm{OS}\leqslant 4-2\sqrt{3}$, which corresponds to
$d\gtrsim 226154704$. Hence it is also not a good choice for achieving
sequential sharing of entanglement.

When the two qudits are passed to Alice$_3$ and Bob$_3$, respectively,
from Eq. \eqref{eq4b-2} one can obtain that, if they can sequentially
sharing the entanglement, then one must have $G_1 \geqslant
G_{1,c}^\mathrm{TS}$, $G_2 \geqslant G_{2,c}^\mathrm{TS}$, and
$(1+F_1^2)(1+F_2^2) \geqslant 4G_{1,c}^\mathrm{OS}$. For the optimal
pointer, one can show that the three inequalities hold simultaneously
when $G_{1,c}^\mathrm{OS}\leqslant 1/2$. As we showed previously,
$G_{1,c}^\mathrm{OS}$ decreases to $1/2$ only when $d\rightarrow\infty$,
indicating that the entanglement cannot be shared by Alice$_3$ and
Bob$_3$. As the optimal pointer gives the highest measurement
precision under the same disturbance, one can conclude that the
entanglement can be sequentially shared by at most two Alices and Bobs
via the two-sided weak measurements.

Next, we consider the scenario TS2. Now, the $2^{n-1}$ possible inputs
for Alice$_n$ and Bob$_n$ can be obtained in the same way as that in
Eq. \eqref{eq4a-7}, with, however, $\mathds{E}^{(k)}$ ($k=0,1,\ldots,n-1$)
being replaced by $\mathds{E}^{(k)}\otimes\mathds{E}^{(k)}$. For
$\rho_{A_n B_n}^{\mathds{E}^{(0)} \ldots \mathds{E}^{(n-1)}}$, using
Eq. \eqref{eq3-6} and after a straightforward but somewhat complex
calculation, one can obtain the probability distribution $p_{x_1 x_2}$ as
\begin{equation}\label{4b-3}
\begin{aligned}
 & p_{x_1 x_2|x\in S}= \frac{1+d_1G_n^2 \prod_{k=0}^{n-1}F_k^{2\delta(\mathds{E}^{(k)},\mathds{X})}}{d^2}, \\
 & p_{x_1 x_2|x \notin S}= \frac{1-G_n^2 \prod_{k=0}^{n-1}F_k^{2\delta(\mathds{E}^{(k)},\mathds{X})}}{d^2},
\end{aligned}
\end{equation}
while the probability $q_{z_1 z_1}$ ($q_{z_1 z_2|z_1\neq z_2}$) can
be obtained directly by replacing $\mathds{X}$ in $p_{x_1 x_2|x\in S}$
($p_{x_1 x_2|x \notin S}$) with $\mathds{Z}$. By combining this with
Eq. \eqref{eq2-4} one can obtain the average uncertainty
$\mathcal{U}^{\mathrm{TS}}_{A_n B_n}$, whose form is similar to
$\mathcal{U}^{\mathrm{OS}}_{A_n B}$ in Eq. \eqref{eq4a-9}, with
however the parameters $G_n$ and $F_k$ being replaced by $G_n^2$
and $F_k^2$, respectively. In the following, we determine the number
of Alices and Bobs sharing the entanglement, and we calculate the
critical $G_{n,c}^{\mathrm{TS}}$ numerically, if any, under the
conditions $G_l=G_{l,c}^{\mathrm{TS}}$ ($l=1,\ldots,n-1$).
Specifically, $G_{n,c}^{\mathrm{TS}}$ can be obtained using
the same method as that for obtaining $G_{n,c}^{\mathrm{OS}}$.
Nonetheless, from the results for scenario TS1, one knows that the
entanglement can always be sequentially shared by two Alices and Bobs.

\begin{figure}
\centering
\resizebox{0.44 \textwidth}{!}{%
\includegraphics{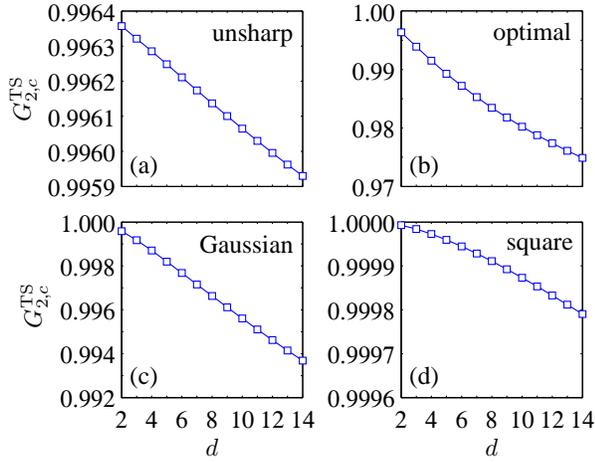}}
\caption{The $d$ dependence of the critical precision
$G_{2,c}^\mathrm{TS}$ (obtained with $G_1=G_{1,c}^\mathrm{TS}$)
for the scenario TS2 with different pointers.} \label{fig:4}
\end{figure}

(B2.1) For the unsharp pointer, $G_{2,c}^{\mathrm{TS}}$ decreases
very slowly with the increase of $d$, see Fig. \ref{fig:4}(a). When
$d\rightarrow\infty$, one can obtain $(G_{2,c}^{\mathrm{TS}})^2
\rightarrow 2(5+2\sqrt{2})/17$. However, when the previous two Alices and
Bobs can sequentially share the entanglement, one always has
$\mathcal{U}^{\mathrm{TS}}_{A_3B_3}>\log_2 d$. Thus, at most, two Alices
and Bobs can sequentially sharing the entanglement in this case.

(B2.2) For the optimal pointer, as shown in Fig. \ref{fig:4}(b),
$G^\mathrm{TS}_{2,c}$ still decreases with the increase of $d$,
and when $d \rightarrow \infty$, one has
$G_{2,c}^\mathrm{TS}\rightarrow \sqrt{6}/3$. Moreover, when the
previous two Alices and Bobs can share the entanglement,
$\mathcal{U}^{\mathrm{TS}}_{A_3B_3}>\log_2 d$. Thus Alice$_3$ and
Bob$_3$ cannot share the entanglement even with the optimal pointer.
This result also indicates that, for the weak measurement with
arbitrary pointer state, the entanglement can be shared by, at most,
two Alices and Bobs.

(B2.3) For the Gaussian pointer, the $d$ dependence of
$G_{2,c}^\mathrm{TS}$ is shown in Fig. \ref{fig:4}(c). When
$d\geqslant 10$, it becomes smaller than that for the unsharp
pointer, indicating that for the two-qudit system with large local
dimension, the Gaussian pointer performs better than the unsharp
pointer on entanglement sharing.

(B2.4) For the square pointer, as illustrated in Fig. \ref{fig:4}(d),
$G_{2,c}^\mathrm{TS}$ is obviously larger than that for the other
pointers. It decreases slowly with the increasing $d$, and when
$d\rightarrow \infty$, it approaches the same asymptotic value as
that with an unsharp pointer due to the same reason for the
one-sided scenario.

\subsection{Case of nonmaximally entangled states} \label{sec:4c}
Up to now, we demonstrated sequential sharing of entanglement
for the initial maximally entangled state, then it is natural to
ask whether the sequential sharing of entanglement is also possible
for those nonmaximally entangled states. To this end, we consider
the following isotropic state:
\begin{equation}\label{eq4c-1}
 \rho_I= p|\Psi^+\rangle\langle\Psi^+| + \frac{1-p}{d^2}\iden,
\end{equation}
where $0\leqslant p\leqslant 1$ and $\rho_I$ is separable for
$p\leqslant 1/(d+1)$. Note that $\rho_I$ in Eq. \eqref{eq4c-1}
is equivalent to that given in Ref. \cite{isotropic}.

For this state, assuming that all other conditions of the
scenarios remain unchanged, then after some algebra one can
obtain the expressions of $U^\alpha_{A_n B}$, $ U^\alpha_{A_n B_n}$,
$\mathcal{U}^\alpha_{A_n B}$, and $\mathcal{U}^\alpha_{A_n B_n}$
($\alpha=\mathrm{OS}$ or $\mathrm{TS}$), which are similar to those
listed in the above two subsections, and the only difference lies
in that the parameters $G_n$ for the one-sided case and $G_n^2$ for
the two-sided case are multiplied by a factor $p$. Then one can show
that for $p<p_1= G_{1,c}^{\mathrm{OS}}$, the entanglement in $\rho_I$
cannot be shared by any observer, irrespective of the measurement
scenario.

For the scenario OS1, from Eq. \eqref{eq4a-4} (note that $G_n$
should be replaced by $pG_n$) one can obtain that if two Alices
can share the entanglement in $\rho_I$ with Bob, then the following
inequalities must hold:
\begin{equation}\label{eq4c-2}
 G_1\geqslant \frac{G_{1,c}^{\mathrm{OS}}}{p},~
 F_1\geqslant \frac{2G_{1,c}^{\mathrm{OS}}}{p}-1.
\end{equation}
Then for the unsharp pointer, they hold simultaneously when
$p\geqslant p_2^\mathrm{OS}$, where
$p_2^\mathrm{OS}=(3d-\sqrt{2d}-4)G_{1,c}^{\mathrm{OS}}/2d_2$.
Similarly, one has $p_2^\mathrm{OS}= 5G_{1,c}^{\mathrm{OS}}/4$
for the optimal pointer and
$p_2^\mathrm{OS}= 3G_{1,c}^{\mathrm{OS}}/2$ for the square pointer.
For the Gaussian pointer, $p_2^\mathrm{OS}$ can be obtained numerically.
They all decrease with the increasing $d$ (for the conciseness of this
paper, we do not plot them here). For the scenario TS1, following
the same line of reasoning as above, one can obtain
$p_2^\mathrm{TS}= 3G_{1,c}^{\mathrm{OS}}/2$ for the optimal
pointer, while for the other three pointers, even for $p=1$, the
entanglement can be shared by two Alices and Bobs only for very
large $d$, so we do not derive the corresponding
$p_2^\mathrm{TS}$ here.

\begin{figure}
\centering
\resizebox{0.44 \textwidth}{!}{%
\includegraphics{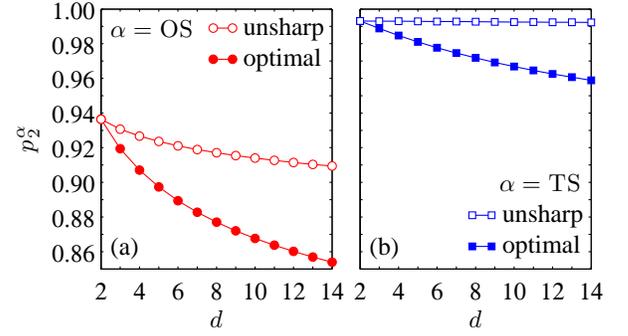}}
\caption{The $d$ dependence of $p_{2}^\alpha$ ($\alpha=\mathrm{OS}$
or $\mathrm{TS}$) for the scenarios OS2 and TS2 with unsharp and
optimal pointers.} \label{fig:5}
\end{figure}

For the scenarios OS2 and TS2, the critical $p_{2}^\alpha$
larger than which the entanglement can be shared by two
Alices and Bob can be obtained numerically. We illustrate
the method for OS2 (it is the same for TS2).
First, we assume $G_1=G_{1,c}^\mathrm{OS}/p$ which
ensures that Alice$_1$ and Bob share the entanglement. For
given $G_1$, the associated $F_1$ can be obtained
accordingly for different pointers. Then by substituting
$G_1$, $F_1$, and $G_2=1$ into Eq. \eqref{eq4a-9},
$\mathcal{U}^{\mathrm{OS}}_{A_2 B}$ will be transformed to
a function of $p$, and by solving numerically the equation
$\mathcal{U}^{\mathrm{OS}}_{A_2 B}=\log_2 d$, we can obtain
the critical $p_{2}^\alpha$.
For the unsharp and optimal pointers, the $d$
dependence of $p_{2}^\alpha$ can be found from Fig. \ref{fig:5}
(the $d$ dependence of $p_{2}^\alpha$ for the square and
Gaussian pointers are similar and we do not plot them here).
They also decrease with the increase of $d$. For infinite
large $d$, one has $p_2^\mathrm{OS} \rightarrow 2/3$ and
$p_2^\mathrm{TS} \rightarrow 2(5+2\sqrt{2})/17$ for the unsharp
pointer, while $p_2^\mathrm{OS}\rightarrow 4-2\sqrt{3}$ and
$p_2^\mathrm{TS} \rightarrow 2/3$ for the optimal pointer.

The above results exemplify that even the state is not maximally
entangled, it is also possible to sequentially sharing the
entanglement in it. From Ref. \cite{isotropic} one can note that
in the region of $p>p_{2}^\alpha$, the entanglement of formation
(a measure of entanglement \cite{EoF}) increases monotonically
with the increase of $p$, thus the condition $p>p_{2}^\alpha$ for
sequential sharing of the nonmaximal entanglement implies that
the entanglement in $\rho_I$ must be stronger than a critical
value.

\subsection{Other cases} \label{sec:4d}
Having clarified the number of observers sharing the entanglement
for $\mathds{X}$ and $\mathds{Z}$ constructed via the MUBs given
in Eq. \eqref{eq3-7}, we now turn to the case of primes $d$, for
which one can also construct $\mathds{X}$ and $\mathds{Z}$ via
other MUBs. For this case, when considering the one-sided
scenario, a direct calculation shows that the uncertainties and
the corresponding critical measurement precision are the same as
those discussed above. When considering the two-sided scenario,
however, they remain the same as that discussed above only for
$d=2$. For primes $d\geqslant 3$, $U^{\mathrm{TS}}_{A_nB_n}$ and
$\mathcal{U}^{\mathrm{TS}}_{A_nB_n}$ obtained with $\mathds{X}$
and $\mathds{Z}$ constructed by other MUBs are always larger than
that obtained with $\mathds{X}$ and $\mathds{Z}$ constructed via
Eq. \eqref{eq3-7}. As a matter of fact, for this case even
Alice$_1$ and Bob$_1$ cannot witness the entanglement.

Another issue that remains is whether the conclusion also holds when
different observers choose different measurement settings for the
primes $d\geqslant 2$, e.g., Alice$_1$ (or Alice$_1$ and Bob$_1$)
chooses $\mathds{X}$ and $\mathds{Z}$ of Eq. \eqref{eq3-7} and the
other Alices (Alices and Bobs) choose that constructed by other MUBs.
For this case, a further calculation shows that for both the one- and
two-sided scenarios, at most, one Alice can share the entanglement
with Bob. Thus the measurement settings $\mathds{X}$ and $\mathds{Z}$
constructed via Eq. \eqref{eq3-7} perform better than the other cases.

\begin{figure}
\centering
\resizebox{0.44 \textwidth}{!}{%
\includegraphics{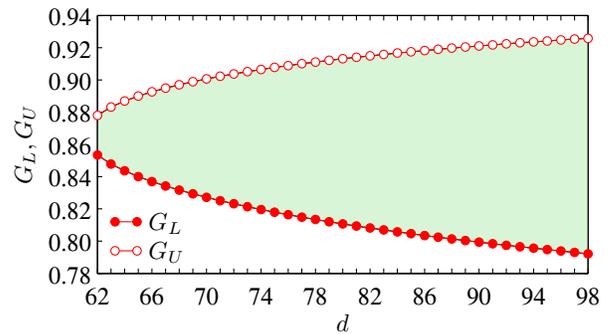}}
\caption{The bounds $G_L$ and $G_U$ versus $d$ for the scenario
OS1 with an optimal pointer. When $G$ locates in the green shaded region, the
maximum entanglement can be shared by two Alices and Bob.} \label{fig:6}
\end{figure}

Finally, it is also relevant to ask whether or not the number of
observers sharing the entanglement will be changed if they
measure the received qudits with equal precision, i.e., $G_n=G$
($\forall n$). For this case, we performed calculations based on
the formulas in the above subsections and it is found that only
for the one-sided weak measurements with the optimal and Gaussian
pointers can there exists a valid region of $G\in [G_L,G_U]$ in
which the entanglement can be shared by two Alices and Bob. For
the scenario OS1 with an optimal pointer,
this region can be obtained by substituting $G_{1,2}=G$
and $F_{1,2}=(1-G^2)^{1/2}$ into Eq. \eqref{eq4a-4}, then the
two bounds $G_L$ and $G_U$, if any, can be obtained by solving
numerically the equation $U^{\mathrm{OS}}_{A_2 B}= \log_2 d$.
They exist for
$d\geqslant 62$ and as illustrated in Fig. \ref{fig:6}, the bound
$G_L$ ($G_U$) decreases (increases) with the increase of $d$.
When $d \rightarrow \infty$, $G_L$ approaches the asymptotic value
of about 0.5437 and $G_U$ approaches the asymptotic value 1. These
asymptotic values are solutions of $G^4-2G+1=0$, which was obtained
from Eq. \eqref{eq4a-4} for the optimal pointer.
For the scenario OS2 with an optimal pointer, the region
$G\in [G_L,G_U]$ can be obtained similarly.
It exists for $d \geqslant 11$, where $G_L$ and $G_U$ approach the
same asymptotic values as above; in particular, $G_U$ approaches
rapidly the asymptotic value 1 (e.g., it is of about 0.9931 for
$d=11$ and 0.9999 for $d=15$). Similarly, for the scenario OS1
(OS2) with the Gaussian pointer, the region $G\in [G_L,G_U]$ exists
for $d\gtrsim 6950$ ($d\gtrsim 640$), and now the asymptotic value
of $G_L$ is about 0.5861, while the asymptotic value of $G_U$ is
still 1. This shows that the sequential sharing of entanglement is
possible even using weak measurements with near-highest precision
(i.e., very strong but not ideal strong measurement). A similar
phenomenon for nonlocality sharing was reported previously
\cite{shareBT3,shareBE3}.

\section{Conclusion} \label{sec:5}
We explored entanglement sharing from the perspective of
entanglement-assisted EUR, and this enabled us to consider a
general two-qudit state which shows more fruitful characteristics
compared with the two-qubit case. We considered two different
scenarios, that is, the scenario in which multiple Alices and a
single Bob share the entangled pair and the scenario in which
multiple Alices and Bobs share the entangled pair. For both the
scenarios, we considered the weak measurements with different
pointers. The results showed that for the unsharp and square
pointers, at most two Alices could sequentially share the
entanglement with one Bob (one-sided scenario) or two Bobs
(two-sided scenario). For the optimal and Gaussian pointers,
however, the maximum number of Alices sequentially sharing the
entanglement with one Bob increases discontinuously with the
increasing dimension $d$ of each qudit for the one-sided scenario,
while the maximum number of Alices and Bobs remains two for the
two-sided scenario. We also obtained the critical measurement
precision higher than which the entanglement could be shared, and
it was found that it always decreases with the increasing $d$ and
approaches a finite asymptotic value when $d$ approaches infinity.

By further considering the isotropic states, we exemplified that
it is also possible to achieve the sequential sharing of nonmaximal
entanglement, provided that it is strong enough. However, for a general
nonmaximally entangled state, further study is still needed to
clarify the interplay between the amount of entanglement and the
number of observers sharing the entanglement. Furthermore, as the
unsharp measurement is optimal (in the sense of achieving the
highest precision for a given disturbance) only for $d=2$, it is
also of fundamental significance to explore what forms the optimal
pointers take for general weak measurements on high-dimensional systems.

\section*{ACKNOWLEDGMENTS}
This work was supported by the National Natural Science Foundation of China (Grant Nos. 12275212 and 11934018),
Shaanxi Fundamental Science Research Project for Mathematics and Physics (Grant No. 22JSY008),
the Strategic Priority Research Program of Chinese Academy of Sciences (Grant No. XDB28000000),
and Beijing Natural Science Foundation (Grant No. Z200009).

\newcommand{\PRL}{Phys. Rev. Lett. }
\newcommand{\RMP}{Rev. Mod. Phys. }
\newcommand{\PRA}{Phys. Rev. A }
\newcommand{\PRB}{Phys. Rev. B }
\newcommand{\PRD}{Phys. Rev. D }
\newcommand{\PRE}{Phys. Rev. E }
\newcommand{\PRX}{Phys. Rev. X }
\newcommand{\NJP}{New J. Phys. }
\newcommand{\JPA}{J. Phys. A }
\newcommand{\PLA}{Phys. Lett. A }
\newcommand{\NP}{Nat. Phys. }
\newcommand{\NC}{Nat. Commun. }
\newcommand{\EPL}{Europhys. Lett. }
\newcommand{\AdP}{Ann. Phys. (Berlin) }
\newcommand{\AoP}{Ann. Phys. (NY) }
\newcommand{\QIP}{Quantum Inf. Process. }
\newcommand{\PR}{Phys. Rep. }
\newcommand{\SR}{Sci. Rep. }
\newcommand{\JMP}{J. Math. Phys. }
\newcommand{\RPP}{Rep. Prog. Phys. }
\newcommand{\PA}{Physica A }
\newcommand{\CMP}{Commun. Math. Phys. }
\newcommand{\SCPMA}{Sci. China-Phys. Mech. Astron. }


\end{document}